# EFECTO HALL DE ESPÍN INVERSO EN PELÍCULAS DE Nb, Mo y Bi POR BOMBEO DE ESPÍN

D. Ley Domínguez[1], J. A. Matutes-Aquino[2]

[1]Universidad Tecnológica de Ciudad Juárez, Av. Universidad Tecnológica # 3051.Col. Lote Bravo II, Cd. Juárez Chihuahua, México, 32695.
[2]Centro de Investigación en Materiales Avanzados, S.C., Miguel de Cervantes 120, Complejo Industrial Chihuahua, Chihuahua, México, 31109.
Autor Corresponsal: david_ley@utcj.edu.mx

**Resumen:** El efecto Hall de espín inverso utilizado para la detección de corrientes de espín fue observado mediante medidas de voltaje en bicapas de metal normal (MN)/metal ferromagnético (MF), utilizando Nb, Mo y Bi como metal normal y Permalloy (Py, $Ni_{81}Fe_{19}$) como metal ferromagnético. La corriente de espín fue generada por el efecto de bombeo de espín con resonancia ferromagnética. Las muestras fueron depositadas por pulverización catódica (sputtering) con magnetrón de corriente continua a temperatura ambiente sobre sustratos de Si (001). Las tres bicapas de Nb/Py, Mo/Py y Bi/Py tuvieron un acoplamiento espín-órbita lo suficientemente grande para poder observar la generación de voltaje por efecto Hall de espín inverso.

**Palabras clave:** Corriente de espín pura, efecto Hall de espín, efecto de bombeo de espín, resonancia ferromagnética.

**Abstract:** The inverse spin Hall effect used for detection of spin currents was observed by voltage measurements in bilayers of normal metal (NM)/ferromagnetic metal (FM), using Nb, Mo and Bi as normal metal and Permalloy (Py, $Ni_{81}Fe_{19}$) as ferromagnetic metal. The spin current was generated by the spin pumping effect with ferromagnetic resonance. The samples were deposited by dc magnetron sputtering at room temperature on Si (001) substrates. The three bilayers of Nb/Py, Mo/Py and Bi/Py had a spin-orbit coupling large enough to observe the voltage generation by spin Hall effect.

## 1. Introducción

La *espintrónica* es una disciplina que estudia una propiedad intrínseca de los electrones que es el espín, para mejorar la eficiencia y velocidad de los dispositivos electrónicos. Este nuevo tipo de electrónica trabaja no solo con la carga de los electrones sino también con su espín. Los dispositivos espintrónicos combinan la microelectrónica convencional con los efectos derivados de la interacción entre el espín del electrón y las propiedades magnéticas del material. El enfoque para utilizar el espín está basado en su alineación (ya sea "arriba" o "abajo") respecto a una referencia (un campo magnético aplicado o la orientación de la magnetización de una película ferromagnética). Los dispositivos operan con cierta cantidad de corriente eléctrica que depende de una manera predecible de la dirección de alineación del espín. La implementación de la alineación de los espines a la electrónica convencional añade a los dispositivos una mayor capacidad y velocidad con bajo consumo de energía en los dispositivos electrónicos. Actualmente los discos duros de computadora y las memorias MRAM funcionan con dispositivos de espintrónica [1-3].

Los aspectos clave en la espintrónica son la generación y detección de corrientes de espín, en la electrónica convencional la orientación del espín del electrón de la corriente eléctrica o corriente de carga está completamente al azar, como se esquematiza en la figura 1 a), en otras palabras, el espín del electrón no tiene ningún rol en los dispositivos electrónicos, por lo que se ignora por completo el espín del electrón. En contraste con la corriente eléctrica, la *corriente de espines polarizados* toma en consideración la orientación del espín pero también puede estar asociada con la corriente de carga, como se ilustra en la figura 1 b), o puede ser solo un flujo de espín sin flujo neto de carga, llamada *corriente de espín pura*, ilustrada en la figura 1 c). La corriente de espín pura se puede entender como un flujo de electrones polarizados solo con espín hacia arriba (spin-up) sumándole a éste un flujo de electrones igual pero polarizados con espín hacia abajo (spin-down) y fluyendo en dirección opuesta, por esta razón no hay un flujo neto de carga, la corriente de espín pura se puede ver como la diferencia entre el flujo de espín arriba y espín abajo, en contraste con la corriente de carga (corriente eléctrica) donde es la suma de los flujo de espín arriba y espín abajo.

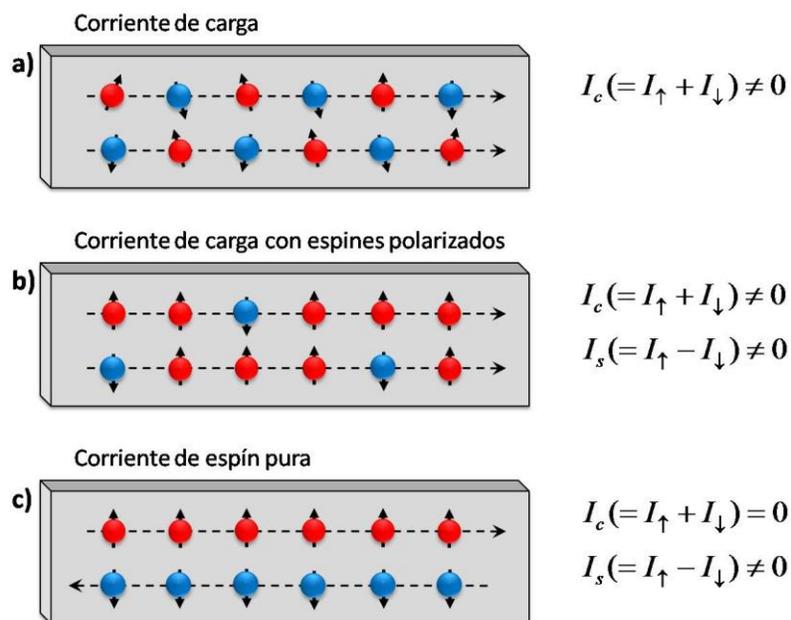

*Figura 1.- Esquema del flujo de electrones: a) corriente eléctrica o corriente de carga, existe un flujo neto de carga y el espín de los electrones está orientado al azar, b) corriente de carga con espines polarizados, tiene un flujo neto de carga y la mayoría de los espines están orientados en la misma dirección, c) No tiene un flujo neto de carga $I_c = I_\uparrow + I_\downarrow = 0$ debido a que $I_\uparrow = -I_\downarrow$ y tiene un flujo neto de espín $I_s = I_\uparrow - I_\downarrow = 2I_\uparrow \neq 0$.*

Las corrientes de espín pueden ser creadas por fenómenos recientemente descubiertos como: (**i**) El efecto Hall de espín (SHE, siglas del inglés *Spin Hall Effect*) como se esquematiza en la figura 2, consiste en una acumulación de espín en los bordes del material en donde existe un flujo de carga eléctrica donde los espines hacia arriba son desviados a un lado y los espines hacia abajo son desviados al otro lado del conductor, si la polaridad de la corriente eléctrica cambia, la orientación de los espines también cambia.

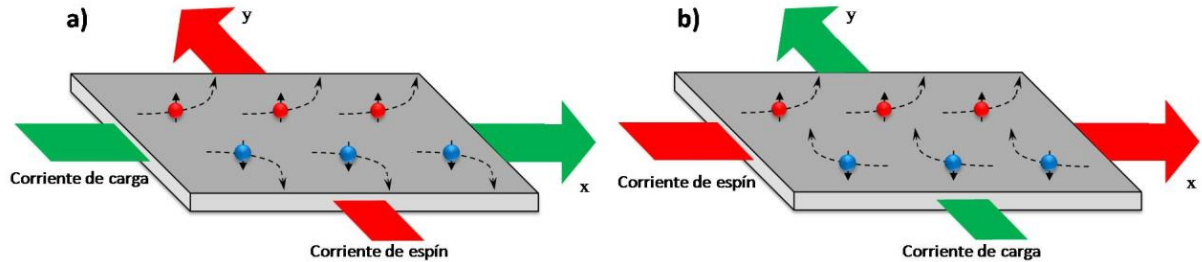

*Figura 2.- Esquema del efecto Hall de espín (SHE) y efecto Hall de espín inverso (ISHE), a) una corriente de carga con espines no polarizados fluye a través de un material no magnético con acoplamiento espín-órbita generando una corriente de espín pura transversal a ésta, b) una corriente de espín pura fluyendo en un material no magnético con acoplamiento espín-órbita genera una corriente eléctrica transversal.*

El efecto Hall de espín en metales no magnéticos (MN) se origina en el acoplamiento espín-órbita, acoplamiento entre el espín del electrón y su movimiento orbital, generalmente encontrado en átomos grandes con un número atómico grande. De manera contraria, si una corriente de espín pura fluye en un material no magnético que tenga fuerte acoplamiento espín-órbita generará un flujo de corriente de carga transversal a la corriente de espín, este efecto es llamado efecto Hall de espín inverso (ISHE, siglas del inglés *Inverse Spin Hall Effect*) [4-8]. (**ii**) El efecto de bombeo de espín (SPE, siglas del inglés Spin Pumping Effect), en donde una corriente de espín pura puede ser inyectada en un material no magnético debido a la precesión de un material ferromagnético adyacente. La precesión de la dirección de la magnetización en el material ferromagnético debido al torque ejercido por un campo de RF (radio frecuencia) externo bombea espines dentro del metal normar adyacente induciendo una corriente de espín pura [9-12]. Los efectos SHE e ISHE son requeridos para la creación de dispositivos espintrónicos siendo el Platino uno de los elementos más utilizado en el estudio de estos efectos [13-14]. Debido al alto costo del platino es importante buscar otras alternativas. En este artículo se reportan medidas de voltaje obtenido del efecto Hall de espín inverso por la corriente de espín generada mediante el efecto de bombeo de espín para los metales de Nb, Mo y Bi.

## 2. Experimentación

Las películas fueron depositadas por la técnica de pulverización catódica (sputtering) con magnetrón de corriente continua a temperatura ambiente, se utilizaron substratos de silicio monocristalino con orientación (001) con un espesor de 500 μm previamente lavados con ultrasonido, se cortaron en rectángulos de 1 mm x 3 mm. Se utilizaran mascaras para depositar los metales normales en el centro de la muestra con una dimensión de 1 mm x 1 mm. La cámara de pulverización fue evacuada a una presión base de $2.2 \times 10^{-7}$ Torr, después fue presurizada con un flujo de argón a $2.4 \times 10^{-3}$ Torr durante el depósito. Se utilizaron blancos de 2 pulgadas de diámetro de Permaloy ($Ni_{81}Fe_{19}$), Niobio, Molibdeno y Bismuto, éstos con una pureza de 99.95%. Todas las bicapas MN/MF de las muestras fueron fabricadas con un espesor de Py de 12 nm y un tiempo de depósito de 1 minuto a 50 mA para las películas de MN. Las muestras fueron puestas en resonancia ferromagnética con un espectrómetro de banda-X operado a una frecuencia de 9.4GHz, la muestra fue posicionada en el centro de una cavidad resonante. Las muestras fueron barridas con un campo magnético externo generado por un electroimán de 9 pulgadas, aplicado en dirección del plano de las películas. El voltaje generado como resultado del efecto Hall de espín inverso se midió con un nanovoltímetro conectado mediante una interface a una computadora. En la figura 3 se observa un esquema de la muestra.

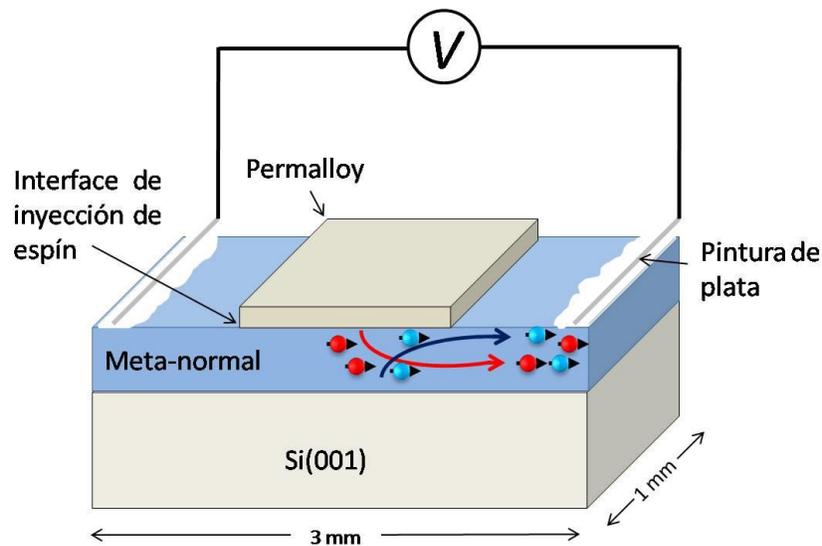

*Figura 3.- Representación de la muestra de interface metal-normal/permalloy.*

## 3. Resultados y discusión

En la figura 4 (a-c) se observa el voltaje detectado por efecto Hall de espín inverso para las muestras de Nb/Py, Mo/Py y Bi/Py. En las tres bicapas se observa un pico de voltaje en aproximadamente 1.1kOe, que es el campo magnético de resonancia del Py para una frecuencia de 9.4GHz.

El pico de voltaje observado para las tres muestras se encuentra en el rango de los µV, valores similares han sido reportados para interfaces de Py/Py [13-14]. El voltaje se genera cuando las bicapas Nb/Py, Mo/Py y Bi/Py están en resonancia ferromagnética, una corriente de espín pura es inyectada de la película ferromagnética a la película de MN por medio del efecto de bombeo de espín. Una vez que existe una corriente de espín pura dentro de la película de MN se genera una acumulación de cargas en las extremidades de la muestra por el desvío de electrones debido al acoplamiento espín-órbita, obteniendo un voltaje de efecto Hall de espín inverso.

La forma del pico de voltaje Hall de espín inverso es la misma que el pico de absorción de resonancia ferromagnética debido a que se crea cuando la película ferromagnética Py se encuentra en un campo magnético y frecuencia adecuada cumpliendo con las condiciones de resonancia, estando el Py en resonancia ferromagnética la inyección de corriente de espín hacia las películas de Nb, Mo y Bi es máxima, obteniendo el máximo voltaje. Una vez saliendo de las condiciones de resonancia el bombeo de espín desaparece, debido a que fuera de las condiciones de resonancia magnética no existe una corriente de espín pura bombeada a la película de MN, el efecto Hall de espín inverso no se acciona y no genera una diferencia de potencial.

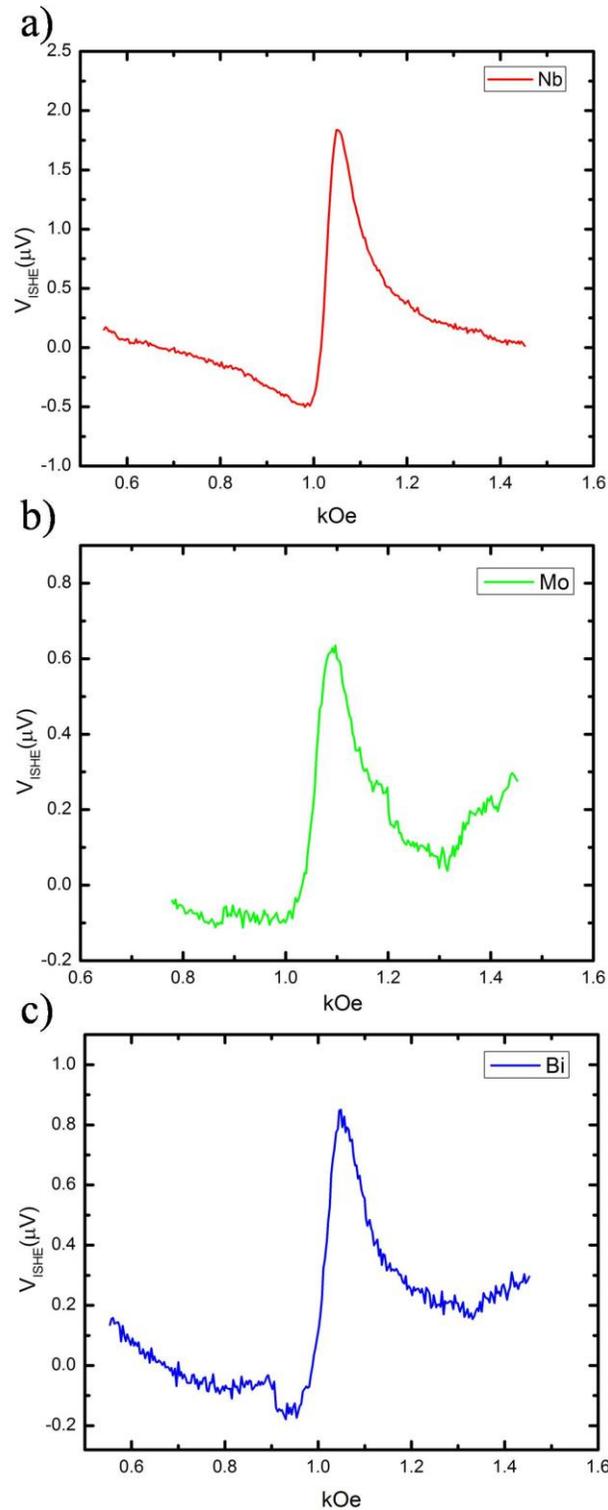

*Figura 4.- Voltaje detectado por efecto Hall de espín inverso (ISHE) generado por la corriente de espín creada mediante el bombeo de espín de resonancia ferromagnética, para las bicapas (a) Nb/Py, (b) Mo/Py y (c) Bi/Py.*

## 4. conclusión

Se obtuvieron medidas de voltaje por efecto Hall de espín inverso en bicapas de Nb/Py, Mo/Py y Bi/Py. Por medio de las medidas de voltaje observadas en los resultados se pudo comprobar que los metales Nb, Mo y Bi tienen un acoplamiento espín-órbita lo suficientemente grande para observar el efecto Hall de espín inverso, por lo que pueden ser utilizados como generadores y detectores de corrientes de espín. Debido al alto costo del Platino, saber que el Nb, Mo y Bi generan voltaje por efecto Hall de espín inverso mediante bombeo de espín es una gran alternativa para minimizar costos en el área de la espintrónica. Estos elementos pueden ser estudiados para desarrollar nuevos dispositivos espintrónicos.